\documentclass[a4paper,fleqn]{cas-dc}

\usepackage[numbers]{natbib}

\def\tsc#1{\csdef{#1}{\textsc{\lowercase{#1}}\xspace}}
\tsc{WGM}
\tsc{QE}
\tsc{EP}
\tsc{PMS}
\tsc{BEC}
\tsc{DE}

\begin{document}
\let\WriteBookmarks\relax
\def\floatpagepagefraction{1}
\def\textpagefraction{.001}
\shorttitle{Leveraging social media news}
\shortauthors{Kepeng Xu et~al.}

\title [mode = title]{Dual Inverse Degradation Network for Real-World SDRTV-to-HDRTV Conversion}                      



\author[1]{Kepeng Xu}[orcid=0000-0003-0650-2442]

\fnmark[1]
\ead{kepengxu11@gmail.com}

\credit{Conceptualization of this study, Methodology, Software}

\affiliation[1]{organization={Xidian University}}

\author[1]{Li Xu}
\fnmark[1]

\author[1]{Gang He}[%
   ]
\cormark[1]
\ead{ghe@xidian.edu.cn}

\credit{Data curation, Writing - Original draft preparation}

\affiliation[2]{organization={Southwest University of Science and Technology}
}

\author[1]{Xianyun Wu}

\author[2]{Zhiqiang Zhang}

\author[2]{Wenxin Yu}

\author[1]{Yunsong Li}



\cortext[cor1]{Corresponding author}
\fntext[fn1]{Kepeng Xu and Li Xu contributed equally to this paper.}

\nonumnote{We first emphasize that existing methods introduce artifacts when converting SDRTV to HDRTV in real-world applications, thus limiting their performance. To address this, we propose DIDNet, which effectively handles inverse tone mapping while suppressing artifact amplification, thereby improving the visual quality of the resulting HDRTV. Key features include an auxiliary loss function for dual degradation task decoupling and a wavelet attention module to enhance frequency details.}

\begin{abstract}
In this study, we address the emerging necessity of converting Standard Dynamic Range Television (SDRTV) content into High Dynamic Range Television (HDRTV) in light of the limited number of native HDRTV content. A principal technical challenge in this conversion is the exacerbation of coding artifacts inherent in SDRTV, which detrimentally impacts the quality of the resulting HDRTV. To address this issue, our method introduces a novel approach that conceptualizes the SDRTV-to-HDRTV conversion as a composite task involving dual degradation restoration. This encompasses inverse tone mapping in conjunction with video restoration. We propose Dual Inversion Downgraded SDRTV to HDRTV Network (DIDNet), which can accurately perform inverse tone mapping while preventing encoding artifacts from being amplified, thereby significantly improving visual quality. DIDNet integrates an intermediate auxiliary loss function to effectively separate the dual degradation restoration tasks and efficient learning of both artifact reduction and inverse tone mapping during end-to-end training. Additionally, DIDNet introduces a spatio-temporal feature alignment module for video frame fusion, which augments texture quality and reduces artifacts. The architecture further includes a dual-modulation convolution mechanism for optimized inverse tone mapping. Recognizing the richer texture and high-frequency information in HDRTV compared to SDRTV, we further introduce a wavelet attention module to enhance frequency features. Our approach demonstrates marked superiority over existing state-of-the-art techniques in terms of quantitative performance and visual quality.

\end{abstract}


\maketitle

\section{Introduction}
\label{Introduction}

\begin{figure*}[htb]
	\includegraphics[width=0.6\linewidth]{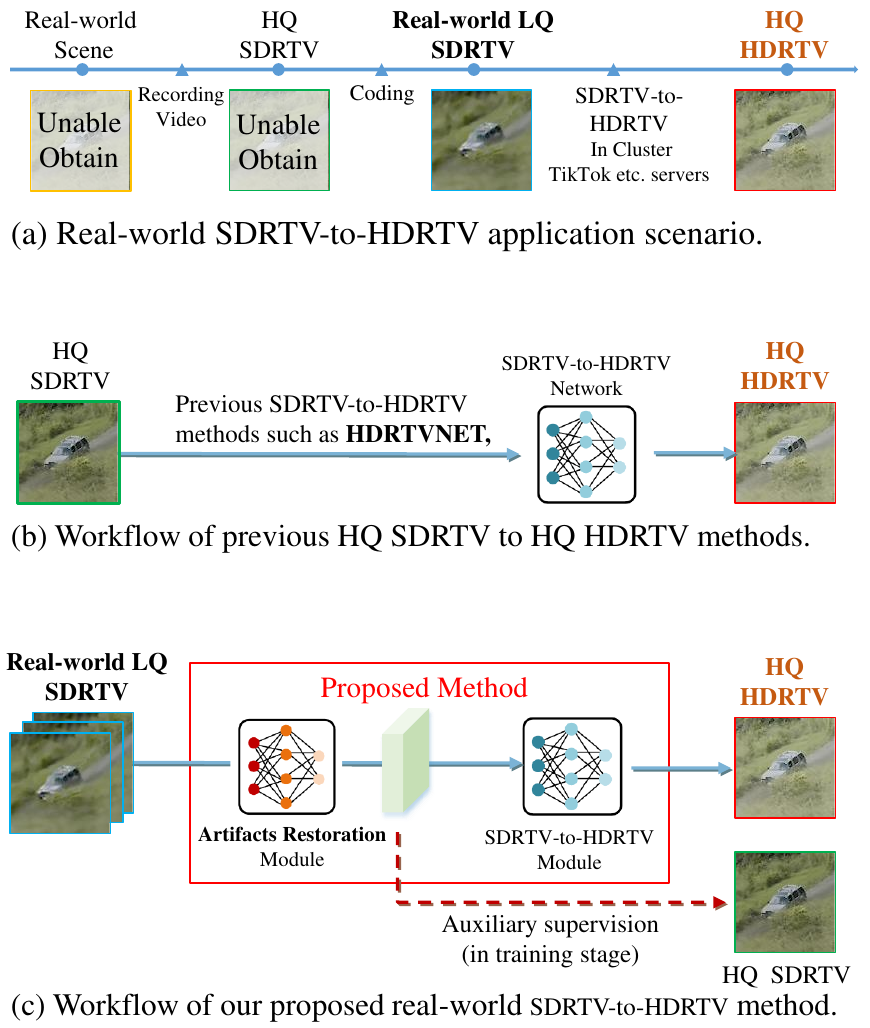}
	\centering
	\caption{
        (a) Real-world SDRTV-to-HDRTV application scenario. Due to distribution copyright and historical technical limitations, media distribution companies hardly obtain high-quality (HQ) SDRTV (nearly lossless) and only possess relatively low-quality (LQ) versions.
        (b) Workflow of previous HQ SDRTV to HQ HDRTV methods. 
        When using these methods with LQ SDRTV in the real world, encoding artifacts are amplified, leading to degraded performance. 
        (c) Workflow of our proposed real-world SDRTV-to-HDRTV method. 
        We propose a dual inverse degradation restoration network to remove encoding artifacts and generate high-fidelity HDRTV results simultaneously.
        }
	\label{exampleframework}       

\end{figure*}

High dynamic range television (HDRTV) has become increasingly popular because it can more realistically reproduce real-world luminance and color information, providing people with a better video viewing experience. The main differences between SDRTV and HDRTV are dynamic range, color gamut, and bit depth. However, despite the advances in HDRTV technology, there is a lack of available HDRTV content. The comparison between HDRTV and SDRTV is shown in Table \ref{hdr_sdr_comparison}. Therefore, the conversion of SDRTV to HDRTV is an important work as it can help to alleviate the scarcity of HDRTV content and improve the video viewing experience for consumers.

Recently, convolutional neural networks are well suited for low-level image and video enhancement. There have been a large number of specific applications, such as image restoration\cite{10420512},image derain\cite{cui2022semi,wang2020dcsfn,wang2022single}, image deblur\cite{licoarse}, image enhancement\cite{ijcai2024p171} and so on. Notably, the CNN-based method has emerged to convert SDRTV to HDRTV.

A large number of researchers have previously introduced deep neural network\cite{ijcai2022p196,xu2024hdrflow} into the field of HDR imaging. The research goal of HDR imaging is to reconstruct irradiation images, which is different from HDRTV technology.
\cite{hdrtvnet, hycondition, 10373884, zhang2023multi}Design different methods to extract SDRTV priors and modulate latent features with the help of prior features, which can convert SDRTV to HDRTV frame by frame.
\cite{fmnet,xu2024dual} proposed a frequency domain processing method to convert SDRTV into HDRTV.
\cite{Guo_2023_CVPR} proposed a method of sub-brightness area processing to process the bright areas and dark areas of SDRTV separately to obtain high-quality HDRTV.
Nonetheless, it must be acknowledged that certain issues remain and require further investigation. Through a comprehensive analysis combining practical requirements for SDRTV to HDRTV conversion with existing technologies, three key observations were identified.

\begin{table*}[ht]
\centering
\begin{tabular}{|c|m{2.5cm}|m{2.5cm}|}
\hline
\textbf{Features} & \centering\textbf{HDR Video} & \centering\textbf{SDR Video} \tabularnewline \hline
Color gamut & \centering Rec. 2020\cite{rec2020} & \centering Rec. 709\cite{rec709} \tabularnewline \hline
Brightness range & \centering $>$ 1000 nits \cite{candelapersquaremetre} & \centering 100 nits \tabularnewline \hline
Color depth & \centering 10-bit & \centering 8-bit \tabularnewline \hline
Contrast ratio & \centering Higher & \centering Lower \tabularnewline \hline
Dynamic range & \centering High & \centering Standard \tabularnewline \hline
Content availability & \centering \textbf{Lack} & \centering Sufficient \tabularnewline \hline
\end{tabular}
\caption{HDRTV VS SDRTV. Compared with SDRTV, HDRTV has obvious advantages in visual quality. However, HDRTV video resources are scarce, so converting SDRTV videos to HDRTV videos is of great significance.}
\label{hdr_sdr_comparison}
\end{table*}

\begin{figure*}
    \centering
    \includegraphics[width=0.8\linewidth]{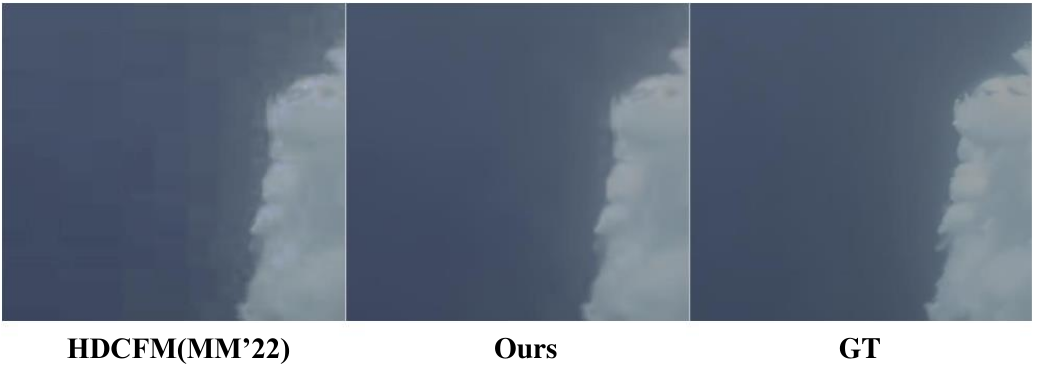}
    \caption{Banding produced by different methods. Previous methods amplify coding artifacts, which can result in subjective quality degradation of the resulting HDRTV.}
    \label{figbanding}
\end{figure*}

The first observation is that \textit{solving the issue of the coding artifacts being amplified during the inverse tone mapping process is indispensable}.
Due to some historical technical and copyright reasons, a large number of current SDRTV videos do not have approximately lossless versions, and only relatively low-quality SDRTVs exist.
Therefore, the SDRTV-to-HDRTV method needs to convert low-quality (LQ) SDRTV to high-quality (HQ) HDRTV in practical applications, as shown in Fig. \ref{exampleframework} (a).
Meanwhile, previous works \cite{dubi1,dubi2} found that the conventional method of converting LQ SDRTV to HDRTV amplifies the coding artifacts. As shown in Fig. \ref{figbanding}, LQ SDRTV with inverse tone mapping exhibits significant coding artifacts. 

The second observation is that the \textit{frame-by-frame SDRTV-to-HDRTV method} extracts the current frame information for feature modulation, and \textit{ignores the multi-frame information for color restoration}. The single-frame method is prone to the discontinuity between frames.

\begin{figure}[b]
    \centering
    \includegraphics[width=0.98\linewidth]{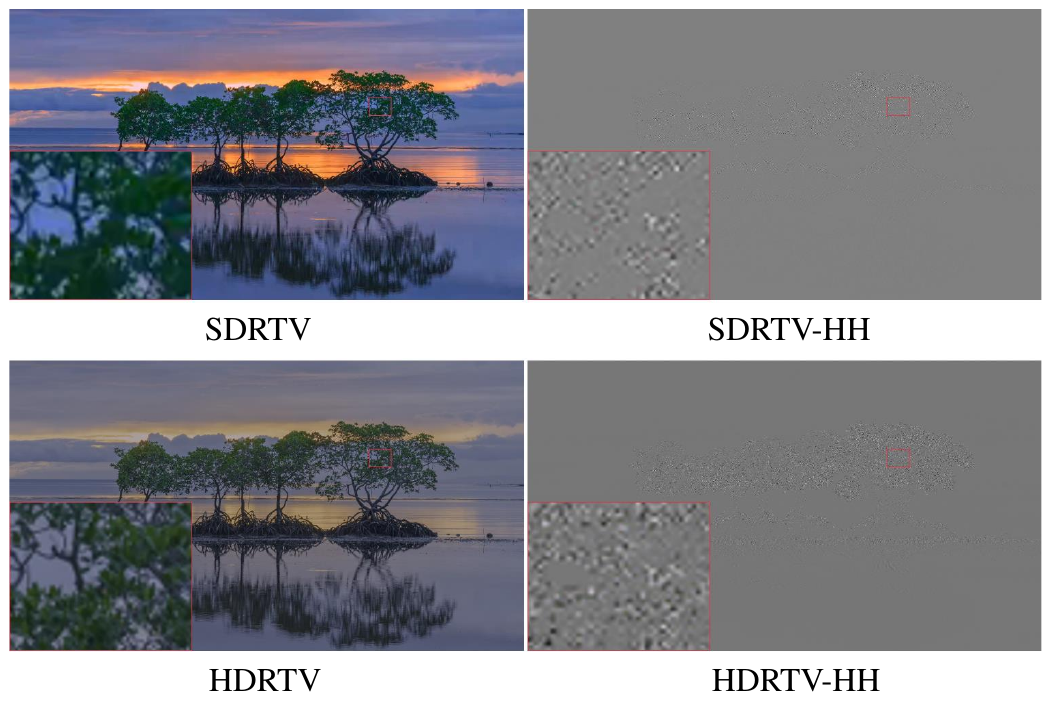}
    \caption{Frequency information comparison. HDRTV has more information in the high frequency area than SDRTV. SDRTV-HH and HDRTV-HH represent the high-frequency subbands obtained by SDRTV and HDRTV through wavelet transform respectively.}
    \label{freqexamp}
\end{figure}

The third observation is that \textit{HDRTV has more information in the high-frequency}, shown as in Fig. \ref{freqexamp}.  Therefore, enhancing features in the frequency domain can effectively improve the quality of HDRTV. 

According to the above three observations, we model practical SDRTV-to-HDRTV as a dual inverse degradation task (video restoration and inverse tone mapping).
Previous methods convert high-quality SDRTV to high-quality HDRTV, as shown in Fig.\ref{exampleframework} (b). But usually SDRTV is not of high quality in real scenes, and this gap will lead to poor performance of such methods.

In our DIDNet, temporal-spatial feature alignment and auxiliary loss are proposed to improve the spatial texture quality of HDRTV.  Furthermore, to improve color restoration quality, a dual modulation convolution that cooperates with a 3D ConditionNet has been designed. Finally, we propose a wavelet attention module to enhance the frequency domain features to further improve the HDRTV quality.
The proposed DIDNet (Fig. \ref{exampleframework} (c)) is able to perform dual degradation recovery simultaneously, making the SDRTV-to-HDRTV method really move towards real applications.

Our contributions can be summarized as four main points:
\begin{itemize}
\item We investigate that the HDRTV obtained by SDRTV-to-HDRTV conversion in real application scenarios has the problem of excessive amplification of coding artifacts. For the first time, a multi-reference frame alignment method is proposed to solve the serious problem of HDRTV artifacts.
\item We reveal inverse tone mapping and artifact restoration are coupled in the process of SDRTV-to-HDRTV. 
Therefore, an auxiliary loss is designed to learn artifact removal, which allows efficient learning of dual restorations using a single end-to-end network.
\item We analyze the computational mode of feature modulation and design a lighter and more efficient dual modulation convolution.
\item We discovered that HDRTV has more high-frequency information, so we proposed wavelet attention to improve the quality of HDRTV in the frequency domain.
\end{itemize}


\section{Related Work}
\label{RelatedWork}

SDRTV-to-HDRTV conversion is the reconstruction of standard dynamic range video (SDRTV) images into high dynamic range video (HDRTV).
Note that SDRTV-to-HDRTV is not the same as HDR imaging, which synthesizes linear irradiation images through multiple frames of LDR images. We briefly introduce some HDR imaging methods.
HDR imaging methods vary, focusing on single-exposure reconstruction, multi-exposure reconstruction, and image quality assessment. Single-exposure techniques\cite{Wang2021DeepLF,LI2024128132,XIAO2024127804,PAN2020147,YAN2017160,XU2024127688,HE2023126590} imitate the LDR formation pipeline using different camera response functions, while multi-exposure methods\cite{9116898} merge several LDR images of varying exposures. Quality assessment\cite{8646579} involves metrics like Peak Signal-to-Noise Ratio, Mean Squared Error, Structural Similarity, and HDR-VDP-2, evaluating HDR images based on criteria like structural differences and visual quality. These diverse approaches highlight the ongoing advancements in HDR imaging research.


\cite{sritm,jsigan} first studied the problem of super-resolution and SDRTV-to-HDRTV.
In these works, the input image is decomposed into a detail component for texture reconstruction and a base component for contrast enhancement.
Specifically, \cite{jsigan} first performs the SDRTV-to-HDRTV conversion using a convolutional neural network (CNN).
Then, \cite{sritm} introduces modulation blocks to modulate the local intensity in a spatially varying manner to achieve adaptive local contrast enhancement.
Recently, \cite{hdrtvnet} has proposed a scheme for SDRTV-to-HDRTV.
Inspired by the SDR/HDR formation process, \cite{hdrtvnet} proposes a three-step solution pipeline that includes adaptive global color mapping, local enhancement, and highlight generation.
\cite{hdrtvnet} also provides a benchmark dataset called HDRTV1K for SDRTV to HDRTV conversion.
\cite{hdcfm,hycondition,HE2023126590} proposes a model for joint local and global feature modulation\cite{fm1,fm2,fm3} capable of local adaptive tuning.
\cite{hdcfm} proposes a feature mapping model and uses dynamic convolution to model feature transformations, thus completing the inverse tone mapping process more accurately.

Although these methods successfully perform inverse tone mapping, the existing SDRTV always has some degradation (coding artifacts).
These artifacts are amplified during the inverse tone mapping process, resulting in poor quality HDRTV from the conversion.
This results in previous methods being ineffective in practical application scenarios.
Unlike the previous methods, this paper addresses the above encoding degradation recovery problem and designs a lighter and more accurate dual modulation convolution for more accurate inverse tone mapping.


\section{Methodology}
\label{Methodology}


\begin{figure*}[ht]
	\includegraphics[width=0.99\linewidth]{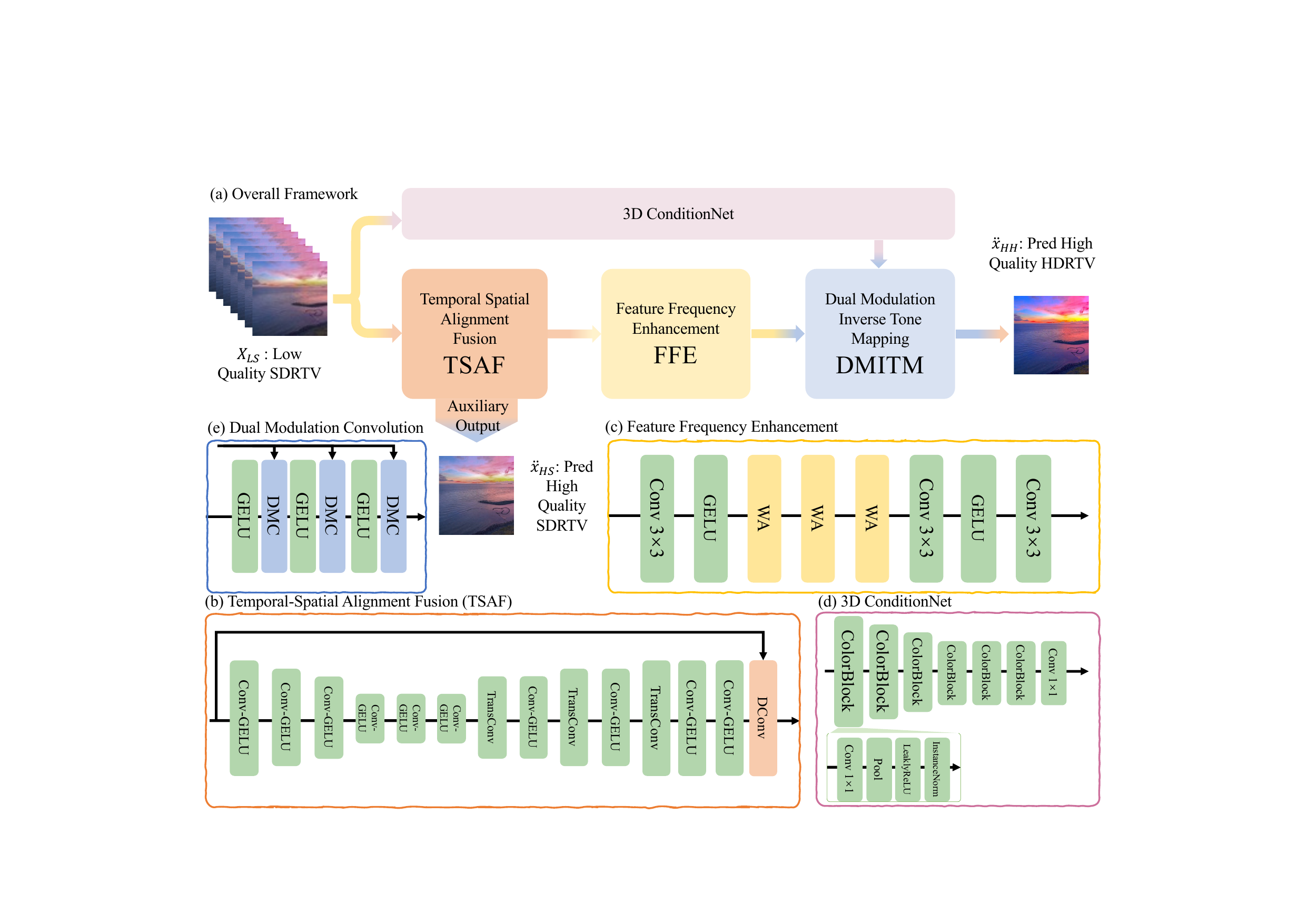}
	\centering
	\caption{
(a) Framework summary. Our framework consists of four parts. (b) Multi-frame alignment artifact repair: Temporal-Spatial Alignment Fusion module \textbf{TSAF}. (c) High-frequency information enhancement: Feature Frequency Enhancement module \textbf{FFE}. (d) Multi-frame color prior extraction: 3D ConditionNet \textbf{3DCN}. (e) Fast color tone mapping: Dual Modulation Inverse Tone Mapping module \textbf{DMITM}. 
   }
	\label{framework}       
\end{figure*}

\subsection{Motivations}

\textbf{Artifacts in Generated HDRTV.}
Typically, video content undergoes a video coding process to reduce storage costs, which results in some artifacts during video coding. The extent of artifact distortion is determined by the bit rate utilized for encoding and the complexity of the scene. In most cases, SDRTV is encoded with an 8-bit depth, leading to the presence of encoding artifacts. As mentioned in \cite{dubi1,dubi2}, these artifacts are amplified during the inverse tone mapping process. If the coding artifacts are not adequately addressed, the visual quality of the resulting HDRTV output will be poor.

\textbf{Limitations of Single-frame Global Feature Modulation (GFM).}
Previous methods only perform inverse tone mapping in the form of single-frame global feature modulation.
Specifically, a state vector is predicted by the image of the current frame, and then global broadcast multiplication and addition is performed on the state vector and the image features extracted from the current frame.
This single-frame adaptive processing is hindered by the lack of continuity between frames. 
Additionally, the computational complexity of feature modulation increases as the video frame resolution increases.

\textbf{Dual Degradation Learning.}
Previous methods only restore from single degradation (tone mapping), but in real-world applications, coding artifacts can lead to the unacceptable visual quality of HDRTV generated by these methods.
SDRTV-to-HDRTV is a dual inverse degradation learning process, i.e., the restoration for coding artifacts and inverse tone mapping. 
This complexity makes it challenging to learn dual degeneration using a single model. A straightforward solution is to employ two separate models where the first model trains only for restoration and the second model learns only inverse tone mapping. However, successive independent training of such multiple submodels leads to cumulative errors and performance degradation due to poor coordination. To address this issue, we propose an auxiliary loss to facilitate coupled learning of dual degradation.

\textbf{Less High Frequency Information in SDRTV.}
During our research, we found that HDRTV contains richer high-frequency information compared to SDRTV, as shown in Figure \ref{figbanding} (b).
Consequently, enhancing the feature in frequency domain can improve the visual quality of HDRTV.

\subsection{Overall of Dual Inverse Degradation Network}

As elaborated by us previously, we propose the dual inverse degradation model to address the issue of coding artifact restoration in converted HDRTV, which allows for efficient learning of dual recovery using a single end-to-end model. The overall framework is presented in Fig.\ref{framework}.

First, the low-quality SDRTV frame $X_{LS}$ is input to the temporal-spatial alignment feature fusion module $TSAF$ to obtain $F_{fusion}$.
At the same time, $X_{LS}$ is input into 3D ConditionNet $3DCN$ to obtain the video color prior $X_P$.
\begin{equation}
   \begin{aligned}
      X_{fusion} & = TSAF(X_{LS}) \\
      X_P        & = 3DCN(X_{LS})
   \end{aligned}
   \label{alignedformual}
\end{equation}
The predicted high-quality SDRTV frames $\ddot{X}_{HS}$ are obtained by convolving $F_{fusion}$ with a $3\times3$ convolution.
\begin{equation}
   \begin{aligned}
      \ddot{X}_{HS} = Conv_{3 \times 3}(X_{fusion})
   \end{aligned}
   \label{conv33formual}
\end{equation}
Auxiliary loss $L_{Aux}$ is computed using $\ddot{X}_{HS}$ with high-quality SDRTV frames $X_{HS}$, thus allowing $TSAF$ to learn the coding artifact inverse degradation process in a targeted manner.
\begin{equation}
   \begin{aligned}
      L_{Aux} = L_1(\ddot{X}_{HS},X_{HS})
   \end{aligned}
   \label{auxlossformual}
\end{equation}
$F_{fusion}$ is input to the frequency domain enhancement module $FFE$, and $F_{FE}$ is obtained after enhancing the frequency domain features.

\begin{equation}
    \begin{aligned}
       X_{FE} & = FFE(X_{fusion}) \\
    \end{aligned}
    \label{outformual1}
\end{equation}

Next, $X_{FE}$ and $X_P$ are input to the dual modulation inverse tone mapping module $DMITM$ to obtain the modulation feature $X_{Modulated}$.
\begin{equation}
    \begin{aligned}
        X_{Modulated} & = DMITM(X_{FE},X_{LS}) \\
        \ddot{X}_{HH} & = Conv 3 \times 3 (X_{Modulated})
    \end{aligned}
     \label{dmcformual1}
\end{equation}

Then $3\times3$ convolution is performed to obtain the predicted high-quality HDRTV frame $\ddot{X}_{HH}$, and the main loss $L_{Main}$ is calculated with the high-quality HDRTV frame $X_{HH}$.
Both loss functions use $L_1$.

\begin{equation}
    \begin{aligned}
       L_{Main} & = L_1(\ddot{X}_{HH},X_{HH})
    \end{aligned}
    \label{outformual}
\end{equation}


\subsection{Restoration: Temporal-Spatial Alignment Fusion Module}



To leverage the temporal information while overcoming artifacts, we propose a temporal-spatial alignment method based on deformable convolution \cite{dcn1, dcn2, dcn3}. The structure of the entire TSAF is introduced in detail in Fig. \ref{framework}. 
The input SDRTV frame $X_{LS}$ is first processed by predicting the offset $F_{offset}$ with an Unet-like structure. Subsequently, deformable convolution is performed using $F_{offset}$ to obtain spatially aligned features $F_{Aligned}$. These aligned features are then further enhanced by aggregating them using residual blocks to obtain the fused features $F_{fusion}$.

The specific definition of deformable convolution is introduced here.  Let \( \mathbf{x} \) be the input feature map, \( \mathbf{y} \) the output feature map, \( \mathbf{k} \) the convolutional kernel, and \( \mathbf{p} \) the standard set of sampling points within the kernel. In standard convolution, the value of \( \mathbf{y} \) at position \( \mathbf{p_0} \) is computed by applying \( \mathbf{k} \) to \( \mathbf{x} \) centered at \( \mathbf{p_0} \). Deformable Convolution extends this by adding a set of offsets \( \Delta\mathbf{p} \), learned by the network, to adjust the positions of each sampling point.

The output of Deformable Convolution, \( \mathbf{y}(\mathbf{p_0}) \), is expressed by:
\begin{equation}
\mathbf{y}(\mathbf{p_0}) = \sum_{\mathbf{p}_n \in \mathbf{P}} \mathbf{k}(\mathbf{p}_n) \cdot \mathbf{x}(\mathbf{p}_0 + \mathbf{p}_n + \Delta \mathbf{p}_n)
\end{equation}

where \( \mathbf{p}_n \) is a standard sampling point in the kernel, and \( \Delta \mathbf{p}_n \) is the corresponding learned offset.

Therefore, deformable convolution can learn the function of multi-frame alignment of pixels in the video, thereby improving the recovery quality.

\subsection{Auxiliary Supervision: Restoring High Quality SDRTV}

In the motivation, it was mentioned that learning coding artifact recovery and inverse tone mapping simultaneously through a single model is challenging due to the coupling of the two degeneracies. 
To address this issue, we propose an auxiliary loss for SDRTV artifact recovery, which directly improves the quality of the SDRTV frames and forces the alignment fusion component to learn the capability of quality enhancement.
 This decoupled learning approach enables a single model to effectively restore both degradations. Specifically, the fused features $F_{fusion}$ are fed into a $3\times3$ convolution to generate predicted high-quality SDRTV frames $\ddot{X}_{HS}$. The training process uses high-quality SDRTV frames $X_{HS}$ as a supervisory signal, which encourages the temporal-spatial alignment fusion module to focus on learning artifact removal and enhancing the SDRTV frames' quality.


\subsection{Feature Frequency Enhancement: Better Restoration of High-Frequency Detail in HDRTV}

Given that High Dynamic Range Television (HDRTV) possesses a greater amount of high-frequency details compared to Standard Dynamic Range Television (SDRTV), our objective was to enhance these details to improve the overall quality of HDRTV. To achieve this, we developed a Feature Frequency Enhancement (FFE) module, focusing on the enhancement of features in the frequency domain. The FFE module employs a wavelet-based attention mechanism, specifically designed to amplify high-frequency components critical for HDRTV.

\begin{figure}[ht]
   \centering
   \includegraphics[width=0.999\linewidth]{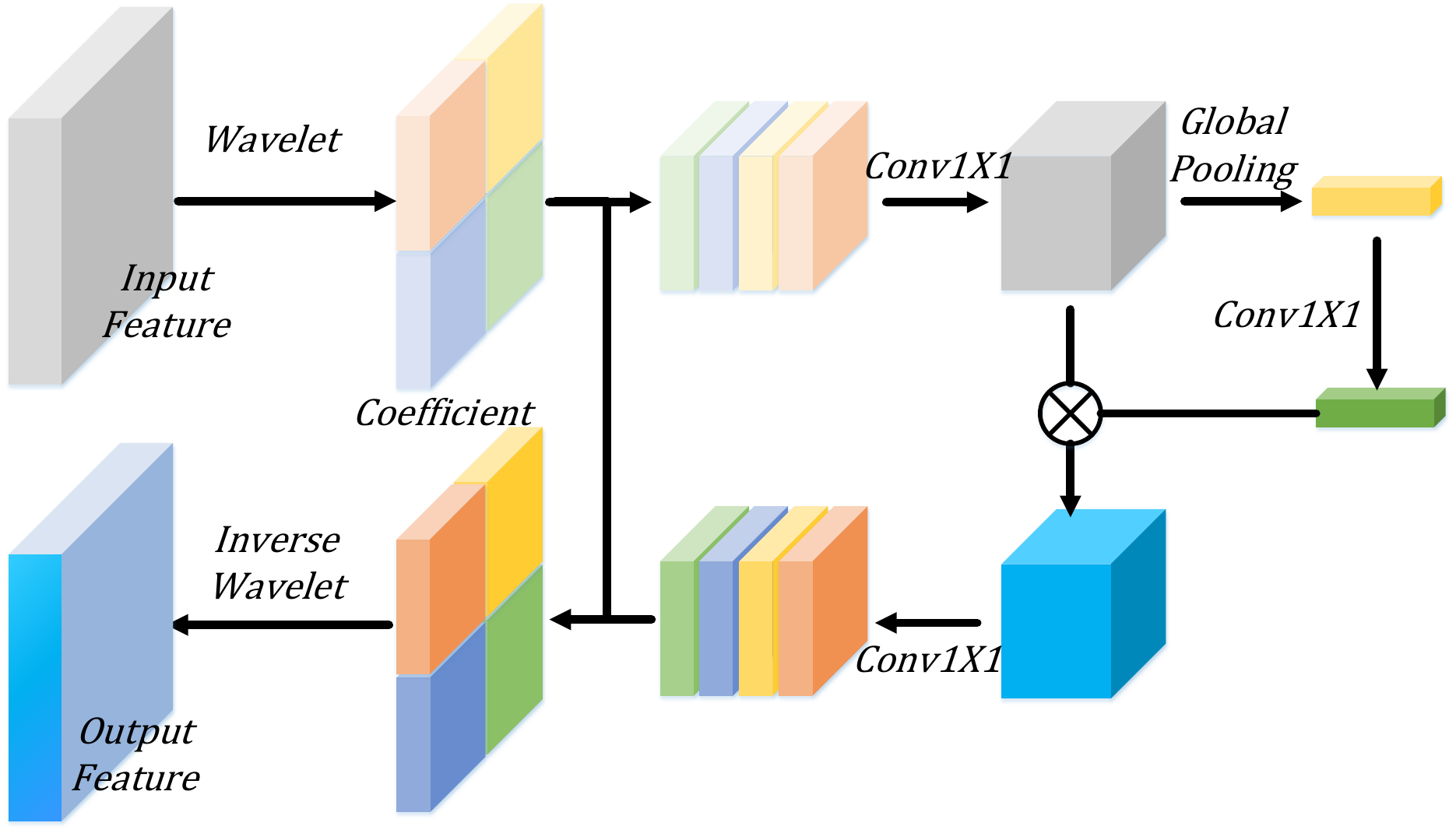}
   \caption{The structure of the Wavelet Attention (WA) module.}
   \label{waveletmodule}
\end{figure}

The mathematical formulation of the FFE module, illustrated in Fig. \ref{waveletmodule}, is as follows:

\begin{equation}
\begin{aligned}
   \mathbf{coeffs} &= \text{DWT}(\mathbf{x}) \\
   \mathbf{z} &= \text{Conv}(\text{Concat}(\mathbf{coeffs})) \\
   \mathbf{s} &= \text{Conv}(\text{GlobalPooling}(\mathbf{z})) \\
   \mathbf{z}_{o} &= \text{Conv}(\mathbf{z} \odot \mathbf{s}) \\
   \mathbf{coeffs}_{o} &= \text{UnConcat}(\mathbf{z}_{o}) \\
   \mathbf{x}_o &= \text{IDWT}(\mathbf{coeffs}_{o} + \mathbf{coeffs})
\end{aligned}
\label{waveletatt}
\end{equation}

The FFE module starts by decomposing the input feature \( \mathbf{x} \) into various frequency subbands \( (ll, lh, hl, hh) \in \mathbf{coeffs} \) using a discrete wavelet transform (DWT). This step is crucial for isolating the high-frequency components, which are more pronounced in HDRTV. Subsequently, these subbands are concatenated and dimensionality-reduced via a \( 1 \times 1 \) convolution to produce \( \mathbf{z} \). Channel attention is then applied to \( \mathbf{z} \), resulting in \( \mathbf{z}_{o} \), which emphasizes the high-frequency details. Finally, the enhanced subbands are recombined and passed through an inverse wavelet transform (IDWT) to construct the output feature \( \mathbf{x}_o \), which exhibits improved high-frequency characteristics while maintaining the integrity of the original image structure.

\subsection{Dual Modulation Inverse Tone Mapping}

\subsubsection{
Preliminary of Global Feature Modulation}
To utilize the global prior extracted from the input image, previous studies \cite{hdrtvnet,hdcfm,hycondition} have introduced a method called Global Feature Modulation (GFM), which has shown to be effective in tasks such as photo retouching and SDRTV-to-HDRTV conversion.
GFM involves modulating the output feature of a convolutional layer using scaling and shifting operations, which is defined in Eq.\ref{gfmformul}.
\begin{equation}
   \begin{aligned}
      y &= \alpha \cdot Conv(x) + \beta 
   \end{aligned}
   \label{gfmformul}
\end{equation}

\noindent
where $\alpha$,$\beta$ are respectively the scaling modulation vector and shift modulation vector required in the feature modulation process. In our case, this vector is predicted by 3D ConditionNet.

\subsubsection{Feature Modulation = Convolutional Kernel Modulation}
GFM modulates the features through scaling and shifting operations using a modulation vector obtained from the condition network. However, we discovered that the same feature scaling and shifting can be achieved by modulating the convolution kernel. To demonstrate this equivalence, we use the example of a $1\times1$ convolution to deduce that modulating features is equivalent to modulating the convolution kernel. This mathematical formal equivalence holds for more complex convolutions, such as $3\times3$ convolutions.


To further clarify this equivalence, we take a specific $1 \times 1$ convolution as an example, assuming that the input feature $\mathbf{x} \in R^{m \times h \times w}$, 
h, w are the row number and columns number of the feature, 
output features $\mathbf{y} \in R^{n \times h \times w}$. Since we use $1 \times 1$ convolution, the processing process for the features at any position in $\mathbf{X}$ is the same. We only consider the features $\mathbf{x_{u,v}} \in R^{m \times 1}$ where the rows and columns are $u,v$. Since $1 \times 1$ convolution is equivalent to the fully connected layer \cite{ma2017equivalence}, we define the convolution kernel of $1 \times 1$ convolution as $\mathbf{W} \in R^{n \times m}$, $bias $ can be defined as $\mathbf{b} \in R^{n \times 1}$. The number of channels of such a convolutional input feature is $m$, and the number of channels of the output feature is $n$. Next, the local operation of general convolution can be defined as $\mathbf{y_{u,v}} = \mathbf{W}\mathbf{x_{u,v}} + \mathbf{b}$.
It should be noted that we perform feature modulation after convolution in this paper. The feature modulation vector is divided into two parts, namely feature scaling and feature shift. The corresponding vectors are defined as $\alpha \in R^{n \times 1}$ and $\mathbf{\beta} \in R^{n \times 1}$. Therefore, the entire process of convolution and feature modulation can be defined in Eq. \ref{fmconv}.

\begin{equation}
\mathbf{y}_{u,v} = \left( \mathbf{W} \mathbf{x}_{u,v} + b \right) \cdot \boldsymbol{\alpha} + \boldsymbol{\beta}
\label{fmconv}
\end{equation}

Next, we can simplify the Eq. \ref{fmconv} as
\begin{equation}
\begin{aligned}
\mathbf{y}_{u,v} &= \mathbf{W}\mathbf{x}_{u,v} \cdot \alpha + \mathbf{b} \cdot \alpha + \boldsymbol{\beta} \\
             &= (\mathbf{W} \cdot \alpha) \mathbf{x}_{u,v} + \mathbf{b} \cdot \alpha + \boldsymbol{\beta}.
\end{aligned}
\label{fmconv2}
\end{equation}

It can be concluded from the Eq. \ref{fmconv2} that the modulation of the feature $\mathbf{x}$ can be transformed into the modulation of the convolution kernel $\mathbf{W}$. Modulating the convolution kernel $\mathbf{W}$ is completely equivalent to the calculation result of the feature $\mathbf{x}$ in mathematical form. However, the amount of calculation required to modulate the convolution kernel is significantly reduced, because the computational cost of modulating the convolution kernel is $n \times m + 2n$, while the cost of feature modulation is $2 \times h \times w \times n $. 
For example, when the input feature is $\mathbf{x} \in R^{64 \times 1080 \times 1920}$ and the output feature is $\mathbf{y} \in R^{64 \times 1080 \times 1920}$. 
The computations of feature modulation is $2 \times 1080 \times 1920 \times 64 = 2.65 \times 10^{8} $, and the computations of the convolution kernel modulation is $64 \times 64 + 128 = 4.22 \times 10^{3} $.
It can be seen that our modulation convolution scheme has a computational advantage of 5 orders of magnitude when processing 1080P size feature.

\subsubsection{Dual Modulated Convolution}

To further enhancing the tone mapping capabilities of our method, the scaling modulation operations are introduced prior to the convolution kernel modulation. This operation incorporates modulation both pre- and post-convolution, thereby boosting the overall effectiveness of our model. 
We call the modulation process as dual modulation convolution (DMC). The definition of dual modulation convolution is as follows:

\begin{equation}
\begin{aligned}
\mathbf{y}_{u,v} &= DMC(\mathbf{x}_{u,v}) \\
             &= (\mathbf{W} \cdot \alpha) (\mathbf{x}_{u,v} \cdot \gamma) + \mathbf{b} \cdot \alpha + \boldsymbol{\beta}
\end{aligned}
\label{dmcv1}
\end{equation}

\noindent
where $\gamma \in R^{m \times 1}$ is the scaling modulation vector of feature modulation before convolution kernel modulation.

Upon analyzing the Eq.\ref{dmcv1}, we can find that the feature modulation process before convolution can still be simplified to the convolution kernel feature modulation, such as the Eq.\ref{dmcv2}.
\begin{equation}
\begin{aligned}
\mathbf{y}_{u,v} &= DMC(\mathbf{x}_{u,v}) \\
                &= (\mathbf{W} \cdot \alpha \cdot \gamma^T) \mathbf{x}_{u,v} + \mathbf{b} \cdot \alpha + \boldsymbol{\beta}
\end{aligned}
\label{dmcv2}
\end{equation}

The proposed dual modulation convolution $DMC$ module has lower computational complexity than ordinary feature modulation. Moreover, it can facilitate the learning of a richer inverse tone mapping process.

\subsection{3D ConditionNet}

Here, we propose a 3D ConditionNet to extract color priors from multiple SDRTV frames. This simple yet effective design pattern can improve the accuracy of feature modulation vector extraction and alleviate inter-frame jitter. The network structure of the 3D ConditionNet is depicted in Figure \ref{framework}(e).

The 3D conditional network is composed of 6 stacked color prior extraction modules $ColorBlock$. Each color prior extraction module is composed of $1 \times 1$ Convolution, Pooling, LeaklyReLU, and InstanceNorm in series. Through a step-by-step downsampling convolution process, the global color prior is extracted.
and our 3d
The conditional network accepts multi-frame SDRTV as input, which brings two advantages:
1. Multi-frame SDRTV can help estimate color priors more accurately.
2. Multi-frame SDRTV color prior extraction can improve the consistency of video multi-frame content.


\section{Experiments}

\begin{figure*}[ht]
	\centering
   \includegraphics[width=0.999\linewidth]{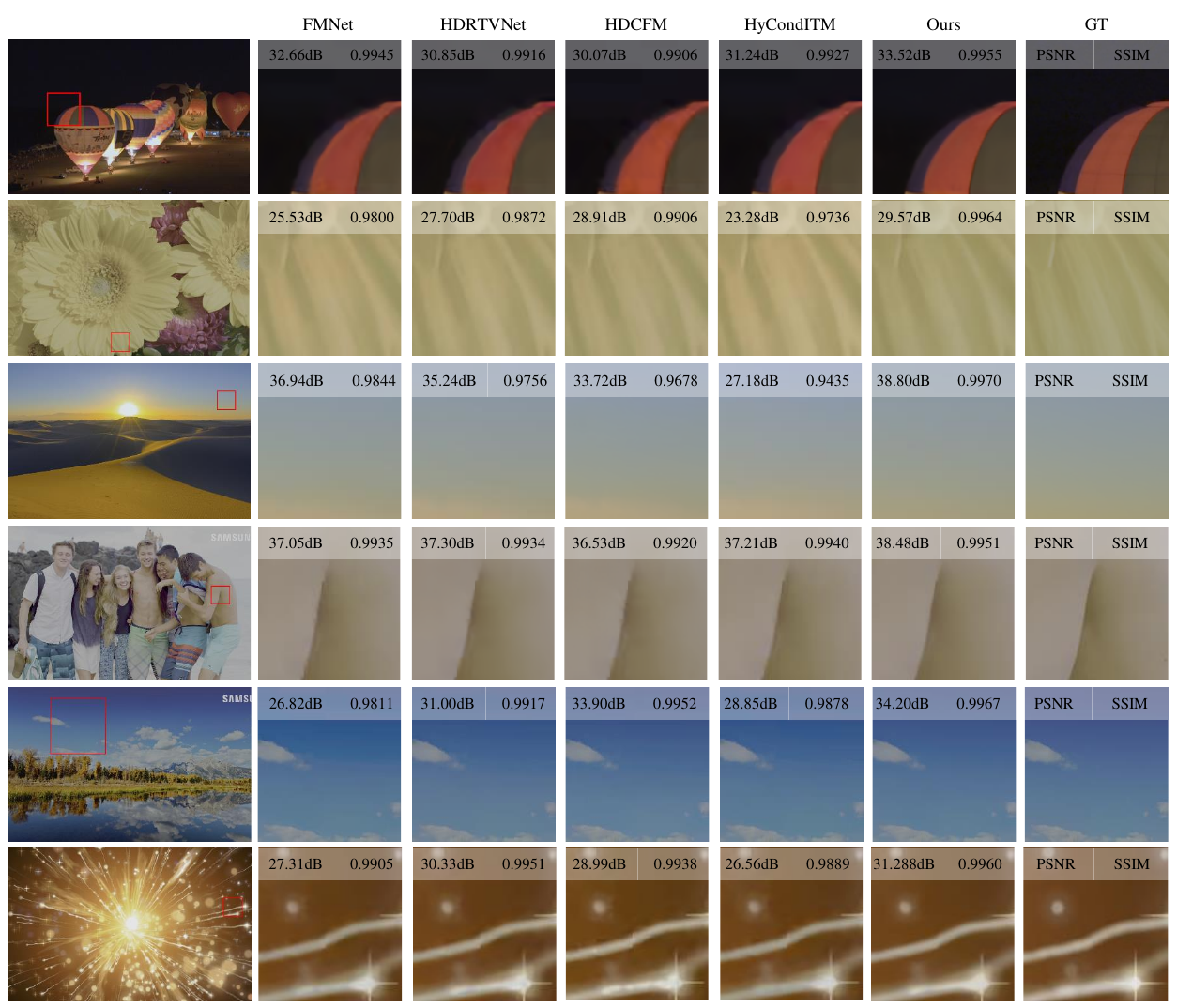}
   \caption{Qualitative results. Our method produces results with fewer artifacts and higher quality than previous methods.}
   \label{resultimage}
\end{figure*}

\subsection{Experiment Settings}

\textbf{Dataset.} Our experiments were based on the HDRTV1K dataset proposed by \cite{hdrtvnet}, which is the first dataset specifically designed for SDRTV-to-HDRTV conversion. This dataset offers several advantages, including a high dynamic range, rich color representation, and complex content. It encompasses a variety of scenes such as urban landscapes, natural settings, and indoor environments. The training set of the HDRTV1K data set contains 18 high-resolution SDRTV-HDRTV video pairs, and the test set contains 4 sets of videos.

For video encoding, we used the widely adopted X265 codec \cite{h265}, known for its high compression efficiency and versatility. Following the approach in previous studies on video codec damage repair, we tested our method under four different Quantization Parameters (QP) values: 27, 32, 37, and 42. This choice was motivated by the need to comprehensively assess the proposed method's ability to convert SDRTV of varying quality, reflecting the uneven quality of videos circulating on the internet. Consistent with the testing methodology in \cite{rrdncnn}, training was conducted on a dataset encoded at QP=32, and testing was performed on SDRTVs encoded at various QP values to evaluate the generalizability of the proposed method.

For experiments setup, we processed the test subset of the HDRTV1K dataset, encoding the SDRTVs in four different QP settings (27, 32, 37, 42), resulting in a total of 348 test image pairs. Each pair consisted of 7 frames of SDRTV and the corresponding middle frame in HDRTV.

\textbf{Implementation Details.} 
We employed the Adam optimizer \cite{adamopt} for our model training, chosen for its efficiency in handling our specific data type, combining the advantages of momentum in gradient descent and the adaptive learning rate feature of RMSprop. The initial learning rate was set to 0.0005, based on optimal learning speed determined from preliminary experiments. To optimize the training process and avoid overfitting, we adopted a strategy where the learning rate is halved every 60,000 iterations after the first 100K iterations. This approach facilitates rapid learning in the initial stages and fine-tuning of the model parameters in the later stages. The total number of training iterations was set to 6.6 million, a figure estimated to be sufficient for the model to reach stable performance levels.

During the training process, we implemented a dual-loss function strategy, assigning a weight of 0.8 to the primary loss and 0.2 to the auxiliary loss. This configuration was designed to ensure that while the model learns to repair artifacts, it does not overlook other crucial visual features. The primary loss focuses on enhancing the overall image quality, while the auxiliary loss concentrates on artifact repair. This dual mechanism allows our model to maintain a balance between artifact repair and overall image quality improvement.

To evaluate the performance of different algorithms comprehensively and fairly, we tested the last six weights stored at the end of training. This method not only considers the model's performance in the latter stages of training but also, by averaging the results of these weights, enables a more accurate assessment of the model's generalization capability across different conditions. This multiple evaluation strategy ensures that our experimental results are not based solely on a single optimal state of the model, but reflect the overall performance of the model under various states.

\textbf{Evaluation Metrics}

The evaluation metrics were carefully selected to comprehensively cover every aspect of video quality assessment. Each metric focuses on a different aspect of video quality, providing a holistic view of the algorithm's performance.

\textbf{PSNR\cite{vvv1} (Peak Signal-to-Noise Ratio):} Widely used to measure image quality, PSNR evaluates the fidelity of an image by comparing the differences between the original and processed images. A higher PSNR value typically indicates lower error and, consequently, higher image quality.

\textbf{SSIM\cite{vvv2} (Structural Similarity Index):} SSIM assesses the structural similarity of images, taking into account changes in luminance, contrast, and structure. An SSIM value close to 1 signifies a high degree of similarity between two images in structure.

\textbf{MS-SSIM\cite{vvv3} (Multi-Scale Structural Similarity Index):} An extension of SSIM, MS-SSIM evaluates structural similarity at various scales. This approach is more aligned with the human visual system's perception of details across different scales.

\textbf{$\Delta E_{ITP}$ \cite{9256392}:} This color fidelity metric assesses the accuracy and authenticity of colors in images by comparing color differences between the original and processed images.

\textbf{VIFp\cite{vifpcite}(Visual Information Fidelity):} VIFp evaluates the visual quality of images by simulating the processing of the human visual system, offering a measure of visual information fidelity.

\textbf{Harrpsi \cite{harrpsicite}:} Focused on local similarity, Harrpsi is particularly suited for assessing the preservation of details and texture information in image regions.

\textbf{VSI\cite{vsicite} (Visual Saliency Index):} VSI accounts for the human visual system's sensitivity to salient features in images, emphasizing the quality of visually significant areas.

These metrics, each evaluating a unique aspect of video quality, collectively provide a comprehensive and detailed assessment of the algorithms' performance in improving video quality.

\subsection{Quantitative Results}

\begin{table}[htbp]
  \centering
  \caption{Performance Comparison of Different SDRTV-to-HDRTV Methods based on PSNR.}
     \resizebox{\linewidth}{!}{%
    \begin{tabular}{cccccc}
    \toprule
    \textbf{Metric(PSNR)} & QP=27 & QP=32 & QP=37 & QP=42 & Mean \\
    \midrule
    CSRNET\cite{he2020conditional} & 33.5977 & 32.4723 & 31.2883 & 29.9460 & 31.8261 \\
    STDF\cite{deng2020spatio}  & 33.9783 & 32.8097 & 31.5913 & 30.1625 & 32.1355 \\
    AILUT\cite{yang2022adaint} & 34.2652 & 33.0583 & 31.7885 & 30.3502 & 32.3655 \\
    AGCM \cite{hdrtvnet} & 34.3363 & 33.1932 & 31.8705 & 30.4065 & 32.4516 \\
    HDRTVNET\cite{hdrtvnet} & 34.2600 & 33.1233 & 31.8782 & 30.3945 & 32.4140 \\
    DEEPSRITM\cite{kim2019deep} & 34.6882 & 33.3317 & 31.9985 & 30.4830 & 32.6253 \\
    FMNet\cite{fmnet} & 34.4618 & 33.4735 & 32.1457 & 30.5838 & 32.6662 \\
    HDRUNET\cite{chen2021hdrunet} & 34.5858 & 33.5913 & 32.2615 & 30.7057 & 32.7861 \\
    HDRTVDM\cite{Guo_2023_CVPR} & 34.5730 & 33.5010 & 32.1828 & 30.7113 & 32.7420 \\
    HDCFM\cite{hdcfm}  & \cellcolor[rgb]{ .996,  .961,  .961}34.8973 & 33.7835 & 32.4403 & 30.9287 & 33.0125 \\
    HyCondITM\cite{hycondition}  & 34.8603 & \cellcolor[rgb]{ .996,  .957,  .957}33.8622 & \cellcolor[rgb]{ .992,  .918,  .918}32.5730 & \cellcolor[rgb]{ .98,  .796,  .796}31.1033 & \cellcolor[rgb]{ .996,  .949,  .949}33.0997 \\
    \midrule
    DIDnet & \cellcolor[rgb]{ .941,  .408,  .408}35.3933 & \cellcolor[rgb]{ .941,  .408,  .408}34.2620 & \cellcolor[rgb]{ .941,  .408,  .408}32.8875 & \cellcolor[rgb]{ .941,  .408,  .408}31.2353 & \cellcolor[rgb]{ .941,  .408,  .408}33.4445 \\
    DIDnet(Tiny) & \cellcolor[rgb]{ .98,  .78,  .78}35.0610 & \cellcolor[rgb]{ .973,  .714,  .714}34.0407 & \cellcolor[rgb]{ .973,  .706,  .706}32.7028 & \cellcolor[rgb]{ .976,  .737,  .737}31.1237 & \cellcolor[rgb]{ .976,  .741,  .741}33.2320 \\
    \bottomrule
    \end{tabular}%
    }
  \label{tabpsnr}%
\end{table}%

\begin{table}[htbp]
  \centering
  \caption{Performance Comparison of Different SDRTV-to-HDRTV Methods based on SSIM.}
     \resizebox{\linewidth}{!}{%
    \begin{tabular}{cccccc}
    \toprule
    \textbf{Metric(SSIM)} & QP=27 & QP=32 & QP=37 & QP=42 & Mean \\
    \midrule
    CSRNET & 0.9756 & 0.9663 & 0.9485 & 0.9169 & 0.9518 \\
    STDF  & 0.9672 & 0.9592 & 0.9429 & 0.9127 & 0.9455 \\
    AILUT & 0.9740 & 0.9645 & 0.9464 & 0.9143 & 0.9498 \\
    AGCM  & \cellcolor[rgb]{ .953,  .49,  .49}0.9773 & 0.9676 & 0.9493 & 0.9172 & 0.9529 \\
    HDRTVNET & 0.9750 & 0.9666 & 0.9499 & 0.9189 & 0.9526 \\
    DEEPSRITM & 0.9750 & 0.9660 & 0.9483 & 0.9168 & 0.9515 \\
    FMNet & 0.9749 & 0.9663 & 0.9495 & 0.9185 & 0.9523 \\
    HDRUNET & 0.9737 & 0.9653 & 0.9488 & 0.9179 & 0.9514 \\
    HDRTVDM & 0.9765 & 0.9680 & 0.9509 & 0.9201 & 0.9539 \\
    HDCFM & \cellcolor[rgb]{ .984,  .82,  .82}0.9772 & 0.9681 & 0.9505 & 0.9194 & 0.9538 \\
    HyCondITM & 0.9770 & \cellcolor[rgb]{ .976,  .729,  .729}0.9691 & \cellcolor[rgb]{ .976,  .741,  .741}0.9530 & \cellcolor[rgb]{ .976,  .741,  .741}0.9227 & \cellcolor[rgb]{ .976,  .745,  .745}0.9554 \\
    \midrule
    DIDnet & \cellcolor[rgb]{ .941,  .408,  .408}0.9773 & \cellcolor[rgb]{ .941,  .408,  .408}0.9697 & \cellcolor[rgb]{ .941,  .408,  .408}0.9540 & \cellcolor[rgb]{ .941,  .408,  .408}0.9240 & \cellcolor[rgb]{ .941,  .408,  .408}0.9563 \\
    DIDnet(Tiny) & 0.9764 & \cellcolor[rgb]{ .988,  .875,  .875}0.9688 & \cellcolor[rgb]{ .973,  .698,  .698}0.9531 & \cellcolor[rgb]{ .961,  .588,  .588}0.9233 & \cellcolor[rgb]{ .976,  .765,  .765}0.9554 \\
    \bottomrule
    \end{tabular}%
    }
  \label{tabssim}%
\end{table}%

\textbf{Results On PSNR}.
Our proposed SDRTV-to-HDRTV network models, DIDNet and DIDNet(Tiny), exhibit a clear advantage in PSNR compared to other methods, showcasing superior results. Detailed results are shown in \textbf{Table \ref{tabpsnr}}. Taking the average PSNR as an example, DIDNet achieves PSNR values of 35.39, 34.26, 32.89, and 31.24 at QP=27, 32, 37, and 42, respectively. Meanwhile, DIDNet(Tiny) achieves PSNR values of 35.06, 34.04, 32.70, and 31.12 at the same QP settings. This indicates that our methods consistently achieve higher quality when converting SDRTV to HDRTV in various compression-quality settings.

Compared to other approaches, DIDNet and DIDNet(Tiny) exhibit outstanding performance across different QP values. This is because our method uses the decoupling strategy, which can effectively remove coding artifacts in the process of learning inverse tone mapping.
And our DIDNet (Tiny) achieves better PSNR results under faster conditions than the past SOTA HyCondITM.

\begin{table}[htbp]
  \centering
  \caption{Performance Comparison of Different SDRTV-to-HDRTV Methods based on MS-SSIM.}
     \resizebox{\linewidth}{!}{%
    \begin{tabular}{cccccc}
    \toprule
    Metric(MS-SSIM) & QP=27 & QP=32 & QP=37 & QP=42 & Mean \\
    \midrule
    CSRNET & 0.9756 & 0.9654 & 0.9470 & 0.9171 & 0.9513 \\

    STDF  & 0.9668 & 0.9578 & 0.9407 & 0.9120 & 0.9443 \\

    AILUT & 0.9740 & 0.9636 & 0.9450 & 0.9151 & 0.9494 \\

    AGCM  & 0.9767 & 0.9662 & 0.9475 & 0.9173 & 0.9519 \\

    HDRTVNET & 0.9753 & 0.9660 & 0.9488 & 0.9199 & 0.9525 \\

    DEEPSRITM & 0.9743 & 0.9645 & 0.9464 & 0.9168 & 0.9505 \\

    FMNet & 0.9747 & 0.9653 & 0.9478 & 0.9185 & 0.9516 \\

    HDRUNET & 0.9743 & 0.9650 & 0.9477 & 0.9184 & 0.9513 \\

    HDRTVDM & \cellcolor[rgb]{ .988,  .875,  .875}0.9774 & 0.9676 & 0.9497 & 0.9202 & 0.9538 \\

    HDCFM & \cellcolor[rgb]{ .941,  .408,  .408}0.9780 & \cellcolor[rgb]{ .98,  .773,  .773}0.9681 & 0.9500 & 0.9204 & 0.9542 \\

    HyCondITM & \cellcolor[rgb]{ .996,  .949,  .949}0.9773 & \cellcolor[rgb]{ .969,  .659,  .659}0.9683 & \cellcolor[rgb]{ .98,  .78,  .78}0.9513 & \cellcolor[rgb]{ .976,  .757,  .757}0.9225 & \cellcolor[rgb]{ .976,  .741,  .741}0.9549 \\
    \midrule
    DIDnet & 0.9773 & \cellcolor[rgb]{ .941,  .408,  .408}0.9687 & \cellcolor[rgb]{ .941,  .408,  .408}0.9521 & \cellcolor[rgb]{ .941,  .408,  .408}0.9237 & \cellcolor[rgb]{ .941,  .408,  .408}0.9554 \\

    DIDnet(Tiny) & 0.9763 & 0.9677 & \cellcolor[rgb]{ .98,  .796,  .796}0.9512 & \cellcolor[rgb]{ .965,  .647,  .647}0.9229 & \cellcolor[rgb]{ .992,  .914,  .914}0.9545 \\
    \bottomrule
    \end{tabular}%
    }
  \label{tabmsssim}%
\end{table}%

\begin{table}[htbp]
  \centering
  \caption{Performance Comparison of Different SDRTV-to-HDRTV Methods based on $\Delta E_{ITP}$.}
     \resizebox{\linewidth}{!}{%
    \begin{tabular}{cccccc}
    \toprule
    Metric(Delta EITP) & QP=27 & QP=32 & QP=37 & QP=42 & Mean \\
    \midrule
    CSRNET & 14.6113 & 16.0855 & 18.0980 & 20.6640 & 17.3647 \\
    STDF  & 14.2508 & 15.4993 & 17.4050 & 19.9490 & 16.7760 \\
    AILUT & 14.1440 & 15.5988 & 17.6174 & 20.1875 & 16.8869 \\
    AGCM  & 12.9657 & 14.5440 & 16.6683 & 19.3322 & 15.8775 \\
    HDRTVNET & 13.0083 & 14.3381 & 16.3008 & 18.9936 & 15.6602 \\
    DEEPSRITM & 13.1191 & 14.5271 & 16.6008 & 19.2724 & 15.8798 \\
    FMNet & 12.3600 & 13.7858 & 15.8880 & 18.6977 & 15.1829 \\
    HDRUNET & \cellcolor[rgb]{ .973,  .757,  .757}11.6940 & \cellcolor[rgb]{ .98,  .812,  .812}13.1480 & \cellcolor[rgb]{ .992,  .867,  .867}15.3129 & \cellcolor[rgb]{ .996,  .945,  .949}18.1809 & \cellcolor[rgb]{ .992,  .843,  .843}14.5840 \\
    HDRTVDM & 12.6855 & 14.1545 & 16.2171 & 18.8657 & 15.4807 \\
    HDCFM & 12.2900 & 13.7847 & 15.8824 & 18.6046 & 15.1404 \\
    HyCondITM & 12.6231 & 13.8869 & 15.8173 & 18.3659 & 15.1733 \\
    \midrule
    DIDnet & \cellcolor[rgb]{ .941,  .408,  .408}11.2686 & \cellcolor[rgb]{ .941,  .408,  .408}12.5932 & \cellcolor[rgb]{ .973,  .412,  .42}14.6117 & \cellcolor[rgb]{ .973,  .412,  .42}17.4066 & \cellcolor[rgb]{ .973,  .412,  .42}13.9700 \\
    DIDnet(Tiny) & \cellcolor[rgb]{ .98,  .831,  .831}11.7824 & \cellcolor[rgb]{ .976,  .776,  .776}13.0999 & \cellcolor[rgb]{ .984,  .737,  .741}15.1126 & \cellcolor[rgb]{ .984,  .737,  .741}17.8797 & \cellcolor[rgb]{ .988,  .761,  .765}14.4687 \\
    \bottomrule
    \end{tabular}%
    }
  \label{tabdeltaeitp}%
\end{table}%

\begin{table*}[htbp]
  \centering
  \caption{Performance Comparison of Different SDRTV-to-HDRTV Methods based on VSI,VIFp and Harrpsi.}
     \resizebox{\linewidth}{!}{%
    \begin{tabular}{cccccc|ccccc|ccccc}
    \toprule
    Method & QP=27 & QP=32 & QP=37 & QP=42 & Mean  & QP=27 & QP=32 & QP=37 & QP=42 & Mean  & QP=27 & QP=32 & QP=37 & QP=42 & Mean \\
    \midrule
    CSRNET & 0.9964 & 0.9956 & 0.9938 & 0.9901 & 0.9940 & \cellcolor[rgb]{ .941,  .408,  .408}0.5855 & \cellcolor[rgb]{ .953,  .525,  .525}0.4973 & 0.3981 & 0.3002 & \cellcolor[rgb]{ .969,  .647,  .647}0.4453 & 0.9126 & 0.8748 & 0.8143 & 0.7316 & 0.8333 \\
    STDF  & 0.9959 & 0.9951 & 0.9935 & 0.9897 & 0.9936 & 0.5584 & 0.4778 & 0.3852 & 0.2921 & 0.4284 & 0.8995 & 0.8653 & 0.8091 & 0.7303 & 0.8260 \\
    AILUT & 0.9964 & 0.9955 & 0.9937 & 0.9896 & 0.9938 & \cellcolor[rgb]{ 1,  .969,  .969}0.5769 & 0.4901 & 0.3924 & 0.2958 & 0.4388 & 0.9132 & 0.8750 & 0.8147 & 0.7321 & 0.8337 \\
    AGCM  & 0.9968 & \cellcolor[rgb]{ .996,  .957,  .957}0.9961 & 0.9942 & 0.9901 & 0.9943 & \cellcolor[rgb]{ .984,  .816,  .816}0.5792 & 0.4918 & 0.3938 & 0.2968 & 0.4404 & \cellcolor[rgb]{ .953,  .518,  .518}0.9206 & 0.8821 & 0.8206 & 0.7368 & 0.8400 \\
    HDRTVNET & 0.9965 & 0.9957 & 0.9940 & 0.9903 & 0.9941 & 0.5620 & 0.4872 & 0.3966 & 0.3025 & 0.4371 & 0.9102 & 0.8757 & 0.8176 & 0.7362 & 0.8349 \\
    DEEPSRITM & 0.9966 & 0.9958 & 0.9940 & 0.9903 & 0.9942 & 0.5732 & 0.4909 & 0.3956 & 0.2995 & 0.4398 & 0.9134 & 0.8773 & 0.8185 & 0.7368 & 0.8365 \\
    FMNet & 0.9966 & 0.9957 & 0.9939 & 0.9902 & 0.9941 & 0.5694 & 0.4921 & 0.3994 & 0.3039 & 0.4412 & 0.9112 & 0.8767 & 0.8192 & 0.7377 & 0.8362 \\
    HDRUNET & 0.9965 & 0.9957 & 0.9939 & 0.9902 & 0.9941 & 0.5644 & 0.4892 & 0.3981 & 0.3028 & 0.4386 & 0.9160 & 0.8812 & 0.8232 & \cellcolor[rgb]{ 1,  .973,  .973}0.7409 & 0.8403 \\
    HDRTVDM & \cellcolor[rgb]{ .941,  .408,  .408}0.9973 & \cellcolor[rgb]{ .941,  .408,  .408}0.9962 & \cellcolor[rgb]{ .961,  .596,  .596}0.9943 & \cellcolor[rgb]{ .98,  .804,  .804}0.9906 & \cellcolor[rgb]{ .941,  .408,  .408}0.9946 & 0.5756 & 0.4910 & 0.3947 & 0.2986 & 0.4400 & \cellcolor[rgb]{ .941,  .408,  .408}0.9208 & \cellcolor[rgb]{ .992,  .914,  .914}0.8834 & 0.8228 & 0.7394 & 0.8416 \\
    HDCFM & 0.9969 & 0.9960 & 0.9942 & 0.9905 & 0.9944 & 0.5755 & 0.4932 & 0.3975 & 0.3006 & 0.4417 & \cellcolor[rgb]{ .957,  .565,  .565}0.9204 & \cellcolor[rgb]{ .996,  .925,  .925}0.8833 & 0.8231 & 0.7399 & \cellcolor[rgb]{ 1,  .984,  .984}0.8417 \\
    HyCondITM & 0.9971 & 0.9959 & 0.9941 & 0.9905 & 0.9944 & 0.5747 & \cellcolor[rgb]{ .941,  .408,  .408}0.4980 & \cellcolor[rgb]{ .941,  .408,  .408}0.4048 & \cellcolor[rgb]{ .973,  .702,  .702}0.3075 & \cellcolor[rgb]{ .941,  .408,  .408}0.4463 & 0.9179 & 0.8828 & \cellcolor[rgb]{ 1,  .965,  .965}0.8237 & 0.7404 & 0.8412 \\
    \midrule
    DIDnet & \cellcolor[rgb]{ .996,  .937,  .937}0.9971 & \cellcolor[rgb]{ .965,  .627,  .627}0.9961 & \cellcolor[rgb]{ .941,  .408,  .408}0.9943 & \cellcolor[rgb]{ .941,  .408,  .408}0.9907 & \cellcolor[rgb]{ .957,  .541,  .541}0.9946 & 0.5660 & 0.4937 & \cellcolor[rgb]{ .953,  .506,  .506}0.4044 & \cellcolor[rgb]{ .949,  .451,  .451}0.3087 & 0.4432 & 0.9176 & \cellcolor[rgb]{ .941,  .408,  .408}0.8842 & \cellcolor[rgb]{ .941,  .408,  .408}0.8269 & \cellcolor[rgb]{ .941,  .408,  .408}0.7446 & \cellcolor[rgb]{ .941,  .408,  .408}0.8433 \\

    DIDnet(Tiny) & \cellcolor[rgb]{ .992,  .886,  .886}0.9971 & 0.9961 & \cellcolor[rgb]{ 1,  .965,  .965}0.9942 & \cellcolor[rgb]{ .961,  .608,  .608}0.9906 & \cellcolor[rgb]{ .98,  .784,  .784}0.9945 & 0.5681 & \cellcolor[rgb]{ .996,  .937,  .937}0.4947 & \cellcolor[rgb]{ .945,  .447,  .447}0.4047 & \cellcolor[rgb]{ .941,  .408,  .408}0.3090 & \cellcolor[rgb]{ .992,  .922,  .922}0.4441 & 0.9167 & 0.8830 & \cellcolor[rgb]{ .965,  .612,  .612}0.8258 & \cellcolor[rgb]{ .957,  .569,  .569}0.7436 & \cellcolor[rgb]{ .98,  .788,  .788}0.8423 \\
    \bottomrule
    \end{tabular}%
    }
  \label{tabvsivifpharrpsi}%
\end{table*}%

\textbf{Results On SSIM and MS-SSIM}.
In the context of SDRTV-to-HDRTV image transformation, our proposed network models, DIDNet and DIDNet(Tiny), have demonstrated a substantial and consistent superiority as quantified by the Structural Similarity Index (SSIM), a critical metric for evaluating image similarity. Through a meticulous analysis of SSIM values across a spectrum of Quantization Parameter (QP) settings, our models consistently outperform their contemporaries, steadfastly upholding heightened image similarity during the intricate conversion process from SDRTV to HDRTV. Notably, at the challenging QP=27 configuration, they achieve markedly elevated SSIM values, thereby establishing the robustness and efficacy of our methodologies. We present a tabulation of experimental results of SSIM evaluation metrics in Table \ref{tabssim}.

Compared with previous methods, DIDNet achieves higher conversion quality at multiple compression quality factors, including the advantage of obtaining higher MS-SSIM scores at high compression quality. Table \ref{tabmsssim} shows the corresponding detailed experimental results.

\textbf{Results on $\Delta E_{ITP}$}.
Our DIDNet model shows significant advantages in the $\Delta E_{ITP}$ (color difference) metric, whose lower value means higher color fidelity. Compared with other models, DIDNet can achieve smaller DeltaEITP values at various compression quality factors (QP), further demonstrating its superior performance in color fidelity. Of particular note is that DIDNet is still able to maintain low DeltaEITP values under high QP values, which highlights its robustness in color fidelity. The experimental results about $\Delta E_{ITP}$ are shown in Table \ref{tabdeltaeitp}.

\textbf{Results On VSI,VIFp and Harrpsi}.
In the context of the Visual Saliency Index (VSI) evaluation, DIDnet demonstrates remarkable superiority across various Compression Quality Factors (QP). Its VSI scores consistently outperform those of its peer models, highlighting DIDnet's adeptness in preserving regions of visual saliency within images. This underscores its outstanding performance in emphasizing the quality of crucial information within images.
When evaluated using the Harrpsi (Local Similarity) metric, DIDnet also exhibits exceptional performance. It consistently achieves high Harrpsi scores across different QP values, indicating its prowess in preserving fine-grained details and texture information within image regions. Particularly, it excels in maintaining local similarity in image regions.
Furthermore, about the VIFp (Visual Information Fidelity) metric, DIDnet once again demonstrates its competitive advantage. Its VIFp scores remain consistently high across various QP values, signifying its proficiency in maintaining the fidelity of visual information within images. This underscores DIDnet's ability to closely mimic the processing mechanisms of the human visual system, further enhancing its performance in preserving image quality.
Detailed experimental results are in Table \ref{tabvsivifpharrpsi}


\subsection{Qualitative Results}

Fig.\ref{resultimage} illustrates the qualitative results obtained from our experimental evaluation of video frames within the realm of SDRTV-to-HDRTV conversion. Our method places distinct emphasis on two critical facets: precise tone mapping and the prevention of pseudo-artifact amplification.

Regarding tone mapping, we discern that our approach markedly enhances the color and luminance characteristics of the converted HDRTV frames. This enhancement is quantitatively validated through subjective evaluations, where viewers can discern and appreciate the notably improved visual fidelity marked by more accurate and visually pleasing color representations.

Furthermore, our method excels in the meticulous prevention of pseudo-artifact amplification. Throughout the inverse tone mapping process, it effectively mitigates the inherent risk of accentuating coding artifacts, thus upholding the overall visual integrity of the converted HDRTV frames. As evident from the visual evidence presented in the figure, our method consistently exhibits significantly fewer artifacts when compared to alternative approaches, resulting in a restored HDRTV output that is characterized by greater clarity and a more natural appearance.

\begin{figure*}[ht]
	\centering
   \includegraphics[width=0.99\linewidth]{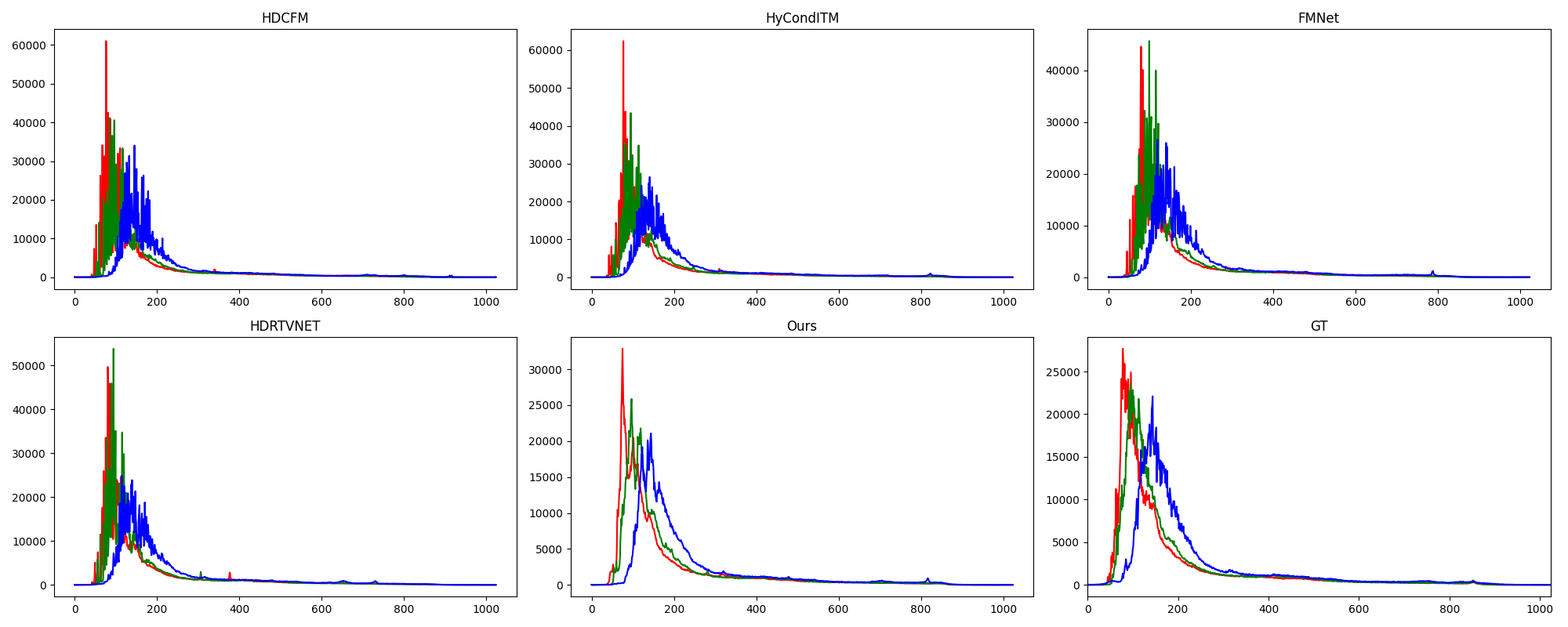}
   \caption{Histogram comparison results. The results output by our method are closer to Ground Truth in the histogram. This is because the proposed method can perform the inverse tone mapping process more accurately, so that the pixel distribution of the output result is more consistent with Ground Truth.}
   \label{hist1}
\end{figure*}

In our study, we conducted histogram analysis of the results generated by various methods and compared them with the histograms of Ground Truth (GT) frames. Through this analysis, we have observed that the histogram of the output generated by our method closely aligns with the histogram of the GT image, indicating that the pixel distribution in the output frames produced by our method is more in line with ground truth. We present the results of the histogram comparison in Fig.\ref{hist1}.

The purpose of this histogram analysis is to validate whether our method accurately captures pixel distribution during the process of generating HDRTV images. Through this analysis, we have identified that our method more precisely represents color distribution, implying that our approach maintains color distribution with greater accuracy when generating HDRTV images, making it closer to the distribution found in real HDRTV images. This result further confirms the exceptional performance of our method in generating high-quality HDRTV images.

\subsection{Ablation Study}
To understand the contributions of the proposed components, we start with a baseline and gradually insert the components. 
The models are trained on QP=37 and the average test results over multiple QPs are reported.

\begin{table}[ht]
   \centering
   \caption{Ablation study on pre-video restoration network, Auxiliary loss (Aux loss), and 3D ConditionNet.}
     \begin{tabular}{lcc}
     \toprule
     Methods & Params (MB) & \multicolumn{1}{l}{ PSNR (dB)} \\
     \midrule
     STDF+HDRTVNet & 2.32 & 32.80 \\
     STDF+HyCondITM & 0.977 & 33.19 \\
     STDF+HDRTVDM  & 0.593 & 32.99 \\
     \midrule
     \textbf{DIDNet} W/o Aux Loss & 0.527 &  33.10 \\
     \textbf{DIDNet} W/o 3DCN & 0.527 & 33.09 \\
     \textbf{DIDNet}& 0.527 & \textbf{33.44} \\
     \bottomrule

     \end{tabular}%
   \label{tableablConditionNet}%
 \end{table}

 \begin{figure*}[ht]
	\centering
   \includegraphics[width=0.99\linewidth]{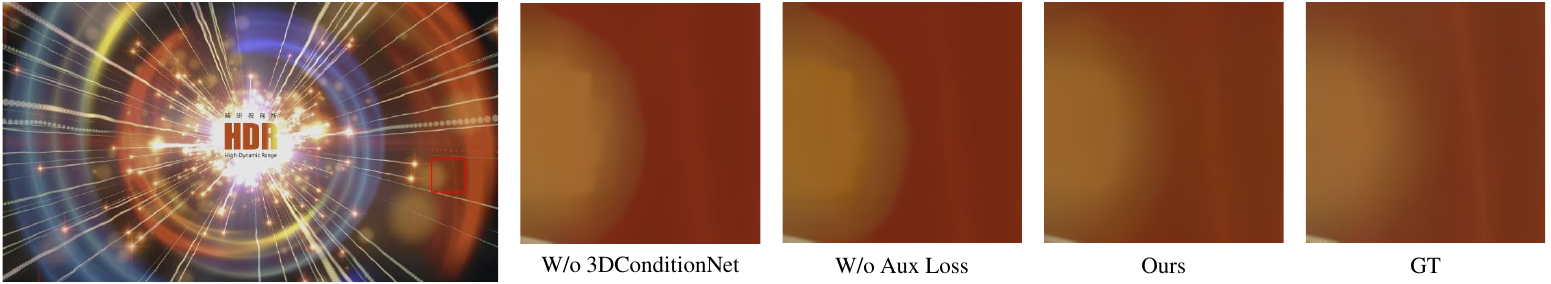}
   \caption{Visual ablation results. The auxiliary loss can greatly reduce coding artifacts, and the 3D conditional network can improve the quality of tone mapping.}
   \label{ablvis1}
\end{figure*}

\begin{table}[ht]
  \centering
  \caption{Detailed ablation results of the wavelet attention module, we calculate the PSNR of the high-frequency part of the wavelet as well as Harrpsi of the model output results, which can evaluate the fidelity of high-frequency information.}
    \begin{tabular}{ccc}
    \toprule
    Ablation WA & HF-PSNR$\uparrow$ & Harrpsi \cite{REISENHOFER201833} $\downarrow$ \\
    \midrule
    W/o WA & 40.03 & 0.8430 \\
    W/  WA & \textbf{40.35} & \textbf{0.8423} \\
    \bottomrule
    \end{tabular}%
  \label{tabwa2}%
\end{table}%

 \begin{table}[b]
   \centering
   \caption{
      Ablation study on dual modulated convolution and wavelet attention in terms of PSNR (dB).
WA and DMC represent Wavelet Attention and Dual Modulated Convolution.
   }
     \begin{tabular}{ccccccc}
     \toprule
     Baseline & WA   &  DMC   & QP27 & QP32 & QP37 & QP42  \\
     \midrule
     $\checkmark$     &                   &                    & 35.24 & 34.12 & 32.75 & 31.13 \\
     $\checkmark$     & $\checkmark$      &                    & 35.30 & 34.23 & 32.88 & 31.25  \\
     $\checkmark$     & $\checkmark$      & $\checkmark$       & 35.39 & 34.26 & 32.89 & 31.24  \\
     \bottomrule
     \end{tabular}%
   \label{ablmain}%
 \end{table}%

 \textbf{Ablation of Video Restoration + Inverse Tone Mapping}.
To verify the effectiveness of the proposed method from components such as auxiliary loss, we connect the inverse tone mapping network with the video restoration network.
Specifically, we designed STDF+HDRTVNet, STDF+ \\HyCondITM, and STDF+HDRVDM.
STDF is currently the most advanced video recovery method (more parameters). The specific experimental results are shown in table\ref{tableablConditionNet}. Compared with STDF+HDRTVNet, STDF+HyCondITM, and STDF+HDRVDM, our method has significant advantages, $\Delta$ PSNR $> 0.25$. This demonstrates that the performance advantage of our approach comes from a decoupled learning approach rather than simple multi-frame alignment.

 \textbf{Ablation of Auxiliary Loss}.
 Since high-quality HDRTV to low-quality SDRTV is a dual degradation process, the model needs to learn both quality enhancement (artifact removal) and inverse tone mapping.
 A single-model learning dual restoration suffers from coupled learning problems, which leads to degraded model performance.
 To solve this problem, we use the auxiliary loss to supervise the dual degradation learning process.
 We show the experimental results of drop auxiliary loss in Table \ref{tableablConditionNet}.
 After dropping the auxiliary loss, the PSNR is dropped by 0.35.
This proves that the intermediate auxiliary loss is crucial for learning the dual degradation process.

 \textbf{Ablation of 3D ConditionNet.}
 To ablate 3D conditional network for prior extraction, we discard multi-frame input conditional network.
 After discarding the 3D conditional, the PSNR drops by 0.36db.
It can be concluded that 3D ConditionNet can estimate color priors more accurately, thereby greatly improving the quality of HDRTV.

 \textbf{Ablation of Wavelet Attention.}
To ablate the Wavelet Attention module, we add the $WA$ module on the basis of the previous module.
 The $WA$ module enhances the features in the frequency domain, which can reconstruct more details and finer edges, and the results are shown in Table \ref{ablmain}. The average PSNR improved from 33.31 to 33.41.
 The gain of $WA$ to high-frequency information is verified by the PSNR of the high-frequency subband of the image. As shown in Table \ref{tabwa2}, after the introduction of $WA$, the high-frequency PSNR increases from 40.03 to 40.35.

\begin{table}[ht]
  \centering
  \caption{Detailed ablation of DMC module. The DMC (Dual Modulation Convolution) module has smaller color difference and higher color fidelity when the calculation amount is much less than the feature modulation.}
    \begin{tabular}{ccc}
    \toprule
    Ablation DMC & $\Delta E_{ITP}$ $\downarrow$  & PSNR $\uparrow$\\
    \midrule
    W/o DMC & 14.31 & 33.41 \\
    W/ DMC & \textbf{13.97} & \textbf{33.44} \\
    \bottomrule
    \end{tabular}%
  \label{tabdmc1}%
\end{table}%

 \begin{table}[ht]
   \centering
   \caption{
   Comparison of the cost of global feature modulation and convolutional kernel modulation.
The computational cost of convolution kernel modulation is lower.
}
     \begin{tabular}{cccc}
     \toprule
     Feature Shape & GFM  &  CKM & Ratio \\
     \midrule
     720$\times$480$\times$64 & 44.22M & 4.22K  & 0.009\% \\
     1080$\times$1920$\times$64 & 265.42M & 4.22K  & 0.001\% \\
     2160$\times$3840$\times$64 & 1061.68M & 4.22K  & 0.0004\% \\
     \bottomrule
     \end{tabular}%
   \label{tableflopsmodulation}%
 \end{table}%

 \textbf{Ablation of  Dual Modulated Convolution.}
 Our motivation for designing DMC is to enhance the tone mapping ability (under the condition of reducing the amount of calculation)
 As shown in Table \ref{tableflopsmodulation}, the cost of modulated convolution is significantly less computationally than feature modulation.
 We conducted ablation, and $DMC$ can improve HDRTV quality.
 It is reported from Table \ref{ablmain} that after adding $DMC$, the PSNR of converted HDRTV is increased by 0.03. As shown in Table \ref{tabdmc1}, after introducing $DMC$, the color difference is reduced from 14.31 to 13.97, the color fidelity is improved.
 This verifies that the proposed $DMC$ module can perform the inverse tone mapping process quickly and accurately.

\subsection{Video Consistency}
 
\begin{table}[htb]
  \centering
  \caption{Experimental results of video consistency. Our method is designed with deformable convolution and 3D conditional convolution, so it can fully utilize the video timing information, so our method has obvious advantages in temporal continuity.}
    \begin{tabular}{ccc}
    \toprule
    Temporal Consistency & std $\Delta E_{ITP}$ $\downarrow$ & MSRL-PSNR $\uparrow$\\
    \midrule
    HyCondITM & 5.45  & 45.63 \\
    \textbf{DIDNet(Ours)} & \textbf{4.64} & \textbf{45.92} \\
    \bottomrule
    \end{tabular}%
  \label{tabts}%
\end{table}%

We calculate the MSRL-PSNR\cite{dai2022video} metric, which can evaluate the video consistency of the proposed method. And we calculated the standard deviation of $\Delta E_{ITP}$, which can evaluate the color fidelity stability of the video sequence.
\cite{dai2022video} proposed a multi-scale continuous loss, and we calculated PSNR based on this loss to obtain MSRL-PSNR.
Table \ref{tabts} shows the experimental results, in which our method outperforms the previous best solutions in terms of video consistency.


\section{Conclusion}

We analyze the difficulties in the current SDRTV-o-HDRTV process, including inaccurate inverse tone mapping, amplified artifacts, difficulties in learning the coupling, and insufficient high-frequency information.
The corresponding modules Dual Modulated Convolution, Auxiliary Loss, and Wavelet Attention are proposed. 
Our proposed method makes SDRTV-to-HDRTV practical, solves the issue of low visual quality caused by the amplification of artifacts, and can convert real-world low quality SDRTV into high-quality HDRTV.
Greatly expanded the application scenarios of SDRTV-to-HDRTV tasks.

\label{Acknowledgment}

\bibliographystyle{cas-model2-names}
\bibliography{cas-refs}




\end{document}